**O. Mostovyi[1], D. Symonov[2]**
[1,2]V. M. Glushkov Institute of Cybernetics of the National Academy of Sciences (NAS) of Ukraine,
Akademika Glushkova Avenue, 40, Kyiv, 03187
[1]alex@mostovyi.net
[2]denys.symonov@gmail.com
[1]https://orcid.org/0009-0006-6687-866X
[2]https://orcid.org/0000-0002-6648-4736

# VULNERABILITY DETECTION IN AARCH64 MACHINE CODE USING A DIGITAL TWIN

**Abstract.** This paper proposes an explainable digital twin for vulnerability detection in AArch64 machine code without access to source code. The digital twin reproduces the concrete execution of a program and preserves the state of registers, processor flags, memory, and live allocated blocks. Each instruction is transformed into a trace event containing the instruction name, operand values, and the post-instruction state. Vulnerabilities are represented as symbolic rules in Kleene algebra with tests: each rule specifies an event sequence and predicates over the machine state. This approach enables the detection of not only isolated unsafe instructions but also multi-step execution patterns. The rules are compiled into finite automata that scan the trace without using an SMT solver. The experimental evaluation covers three CWE classes: integer overflow (CWE-190), null pointer dereference (CWE-476), and heap buffer overflow (CWE-122). The system detected all three predefined vulnerabilities and produced no report on the safe trace. Each detection result includes the triggered rule, the trace position, and the concrete state values, thereby providing a reproducible explanation.

## Introduction

Software-intensive systems increasingly depend on embedded, mobile, and cyber-physical components whose behaviour is determined by low-level executable code. In many practical cases, especially for firmware, third-party libraries, vendor-specific modules, and closed embedded platforms, the source code is unavailable. Vulnerability detection then cannot rely on source-level abstractions such as data types, function boundaries, or high-level control structures. The observable object is the binary execution itself, represented by processor instructions, operands, registers, condition flags, memory addresses, and state transitions.

This setting is particularly important for AArch64, the 64-bit ARM architecture widely used in mobile devices, embedded boards, and edge systems [1]. Security defects at this level may lead to unauthorized memory access, program crashes, privilege escalation, or control-flow manipulation. Binary-level vulnerability detection is therefore both a practical engineering task and a problem of computer science, in which program behavior must be analyzed through formal models of execution, state, and event sequences.

A central difficulty is that many vulnerabilities are not isolated instruction-level defects. They appear as chains of semantically related events. For example, an arithmetic operation may overflow, the corrupted result may later be used as a size parameter, and a subsequent memory operation may access an invalid region. Similar multi-step mechanisms occur in integer overflow, null pointer dereference, heap buffer overflow, use-after-free, and time-of-check to time-of-use scenarios. The Common Weakness Enumeration (CWE) provides a standardized taxonomy for such classes [2], including integer overflow or wraparound (CWE-190) [3], null pointer dereference (CWE-476) [4], and heap-based buffer overflow (CWE-122) [5].

Existing binary analysis methods provide substantial capabilities but do not remove this difficulty. Symbolic execution systems such as KLEE, angr, and BAP analyse program paths and reason about vulnerability-relevant constraints [6]–[8], and SMT solvers such as Z3 with SMT-LIB representations give a strong formal basis for such analysis [9], [10]. These methods may become computationally expensive because of path explosion, complex memory modelling, and architecture-specific instruction semantics. Concrete dynamic analysis is more direct and cheaper, but it observes only executed traces and therefore needs a mechanism for detecting semantically meaningful vulnerability patterns in those traces.

This paper proposes such a mechanism by combining a digital twin of machine-code execution with symbolic artificial intelligence. A digital twin is a virtual representation of a real system or process [11], [12]; here it represents the concrete execution of AArch64 code, recording the machine state after each instruction. The artificial-intelligence component is symbolic and knowledge-based: vulnerability classes are encoded as explicit rules over event sequences, expressed in Kleene algebra, which describes event sequences [13], and in Kleene algebra with tests, which adds state-dependent predicates [14]. A rule can thus require not only that certain instructions occur, but also that concrete conditions over registers, flags, addresses, or memory blocks hold at the corresponding steps. Rule matching is performed by compiling each pattern into a finite automaton and scanning the trace [15].

The unresolved problem addressed here is the lack of a compact, explainable method that operates directly on machine code, detects multi-step vulnerability patterns in concrete traces, does not require an SMT solver during matching, and provides a reproducible data of each detection. The goal of this paper is to develop such a method as a symbolic AI digital twin for selected CWE classes in AArch64 traces. The main contributions are: (i) an AArch64 digital-twin model that represents binary execution as a sequence of state-enriched events; (ii) an encoding of vulnerability knowledge as rules in Kleene algebra with tests; (iii) a solver-free, automaton-based detection procedure over concrete traces; and (iv) an evaluation on traces for CWE-190, CWE-476, and CWE-122, together with a safe trace that verifies no rule fires when the vulnerability conditions are absent.

## Related Work

Automated vulnerability detection in binary code is difficult because the analyzed object is not a source-level program but a sequence of architecture-specific instructions and machine-state transitions. High-level types, variable names, and semantic annotations are unavailable, so the analysis must rely on instruction semantics, register values, condition flags, memory addresses, allocation states, and execution traces. This makes the problem especially relevant for AArch64 binaries used in embedded, mobile, edge, and firmware systems [1].

Existing methods address the problem from several directions. Static binary analysis reconstructs control-flow graphs, lifts machine code into intermediate representations, and applies data-flow or taint analysis; its precision is limited by indirect jumps, compiler optimizations, stripped symbols, and incomplete recovery of memory semantics. Dynamic binary analysis observes actual execution with concrete values; its advantage is precision on the executed path, and its limitation is trace dependence.

Symbolic execution and constraint-based analysis form a more formal class. Tools such as KLEE, Angr, and BAP represent inputs or selected state components symbolically and generate path constraints [6]–[8], which are discharged by SMT solvers such as Z3, often through SMT-LIB [9], [10]. Symbolic execution can reason about alternative paths and synthesize triggering inputs, but its use on real binaries is constrained by path explosion, memory-modelling complexity, and solving overhead, which matter most when the goal is fast analysis of concrete traces rather than exhaustive exploration.

Artificial-intelligence methods form a separate direction. Learning-based detectors built on deep neural networks and graph neural networks can flag vulnerability-prone code fragments, but their output is often probabilistic and does not pinpoint a formally verifiable execution condition. Knowledge-based and optimization methods, by contrast, represent decision criteria explicitly [16], [17]. For binary-level cybersecurity the explicit route is attractive, because it is necessary to know which instruction, flag, address, or event sequence caused the detection. This motivates symbolic artificial intelligence, in which knowledge is represented through rules, predicates, and automata, so that a detection is justified by a matched event pattern and by the truth of state predicates at specific steps [2]–[5].

The unresolved part of the problem is the lack of a compact, understandable system that combines all of the following: direct operation on machine-code traces, explicit symbolic vulnerability rules, concrete AArch64 states, detection of multi-step patterns, and reproducible explanation without invoking an SMT solver per match. Existing static, dynamic, symbolic, and AI-based approaches meet these requirements only partially. The proposed approach integrates an AArch64 digital twin, Kleene Algebra, and automaton-based inference into a modular framework.

**Problem Statement**

Let $P$ be an AArch64 binary program or a finite fragment of machine code, and let $\Sigma$ denote the set of AArch64 instructions supported by the digital twin. Execution is represented through concrete machine states that include registers, processor flags, memory, and live memory blocks. The machine-state space is

$$S = S_{\text{reg}} \times S_{\text{flag}} \times S_{\text{mem}} \times S_{\text{alloc}}, \qquad (1)$$

where $S_{\text{reg}}$ is the space of register valuations, $S_{\text{flag}}$ contains the processor flags $N, Z, C, V$, $S_{\text{mem}}$ represents memory contents, and $S_{\text{alloc}}$ stores the current set of live allocated blocks.

For each step $t = 0, \ldots, T$, let $s_t \in S$ be the machine state after the $t$-th instruction. The corresponding trace event is $e_t = (\iota_t, op_t, s_t)$, where $\iota_t \in \Sigma$ is the executed instruction, $op_t$ is the tuple of operand values, and $s_t$ is the post-instruction state. A concrete execution trace is $\tau = (e_0, e_1, \ldots, e_T)$.

Let $\mathcal{R}$ be a finite knowledge base of vulnerability rules. Each rule $r \in \mathcal{R}$ is a term of Kleene algebra with tests, generated by the grammar

$$r ::= a \mid [\varphi] \mid r_1; r_2 \mid r_1 + r_2 \mid r^*, \qquad (2)$$

where $a$ is an event predicate (with the special predicate any matching every event), $[\varphi]$ is a test over the current machine state, ; denotes sequential composition, $+$ alternative choice, and $(\cdot)^*$ finite repetition. Each test is a predicate $\varphi: S \to \{0,1\}$. Typical tests used for vulnerability detection are

$$V = 1, addr = 0, \neg\, \text{in_alloc}(addr), \qquad (3)$$

corresponding, respectively, to arithmetic overflow, null pointer access, and memory access outside live allocated blocks.

The detection problem is then: given a trace $\tau$ and a rule base $\mathcal{R}$, find all pairs $(r, I)$, $r \in \mathcal{R}$, with $I = (i_0, \ldots, i_k)$ and $0 \le i_0 < \cdots < i_k \le T$, such that the subsequence $\tau_I = (e_{i_0}, \ldots, e_{i_k})$ matches the event structure of $r$ and every test in $r$ is true on the corresponding state $s_{i_j}$. Thus binary-level vulnerability detection is reduced to trace pattern recognition under state-dependent predicates: a detection is valid only when the required event pattern is present and the associated predicates over concrete states hold.

**Materials and methods**

**Architecture of the Intelligent Digital Twin**

The system is organized as an explainable Symbolic AI digital twin for AArch64 vulnerability detection. Its processing pipeline is

$$P \to \text{execution twin} \to \tau \to \text{KAT-based matcher} \to \text{detection report},$$

with four modules: execution twin, event abstraction layer, symbolic knowledge base, and inference and explanation engine. Figure 1 shows the overall flow.

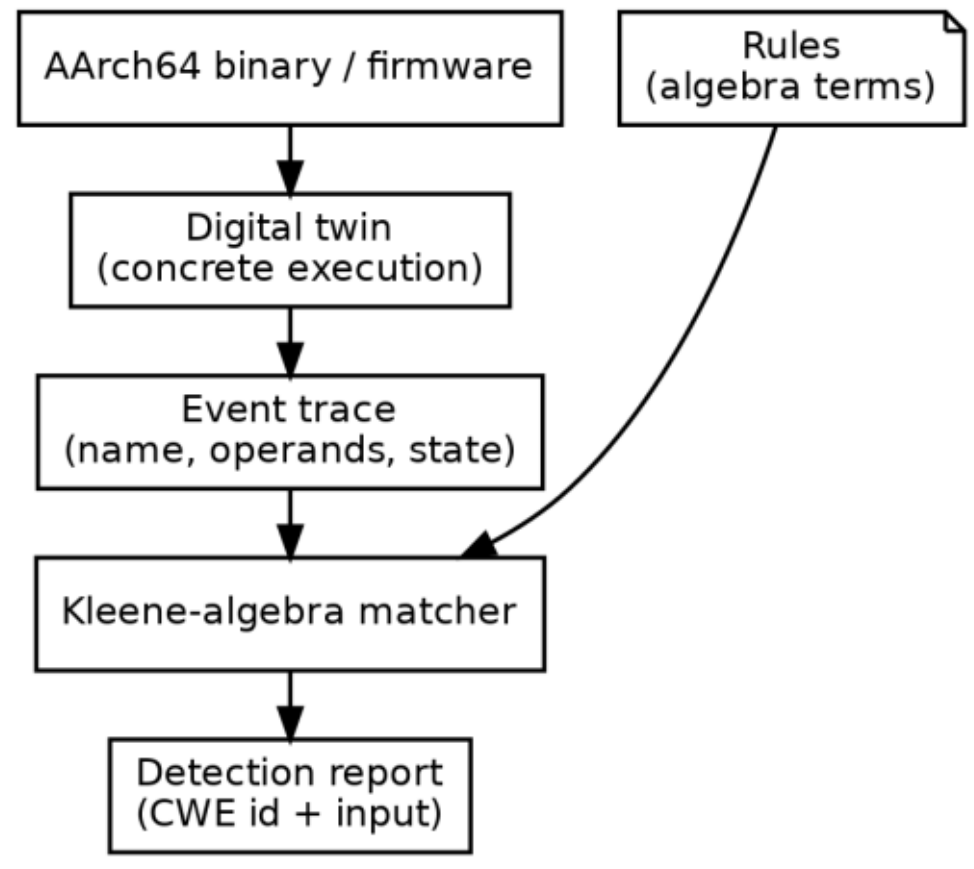

Fig. 1. From a binary program to a detection report

The execution twin reproduces the concrete execution of AArch64 instructions. For each instruction $\iota_t \in \Sigma$ it applies the corresponding semantics and updates the machine state

$$s_t = (reg_t, flag_t, mem_t, alloc_t) \in S, \qquad (4)$$

so that instruction execution is a transition operator $\Delta_\iota: S \times Op_\iota \to S$ with $s_t = \Delta_{\iota_t}(s_{t-1}, op_t)$ . The event abstraction layer then forms the event $e_t = (\iota_t, op_t, s_t)$ defined in the Problem Statement, so the run becomes a state-enriched trace $\tau = (e_0, \dots, e_T)$ rather than raw bytes or disassembled text.

The symbolic knowledge base holds a finite set $\mathcal{R}$ of vulnerability rules. Each rule combines an event pattern with predicates over machine states and is stored as a Kleene-algebra-with-tests term or an equivalent JSON description. This module is the component of Symbolic AI: vulnerability knowledge is explicit, interpretable, and extendable.

The inference and explanation engine compiles each rule $r$ into an automaton

$$A_r = (Q_r, q_r^0, F_r, \delta_r), \qquad (5)$$

where $Q_r$ is the set of states, $q_r^0$ the initial state, $F_r$ the accepting states, and $\delta_r$ the transition relation induced by the event pattern and tests of $r$. The engine scans $\tau$ sequentially; at each event it checks both the instruction pattern and the relevant predicate $\varphi(s_t)$, and reports a vulnerability only when the required sequence is matched and all tests are true. A detection has the form

$$d = (r, \text{CWE}, I, V_d), \qquad (6)$$

where $r$ is the triggered rule, CWE the weakness identifier, $I = (i_0, \dots, i_k)$ the matched trace positions, and $V_d$ the concrete values that justify the detection (for example $V = 1$, $addr = 0$, a destination register, or a store address outside all live blocks). The architecture thus separates execution modelling, event abstraction, symbolic knowledge, and inference, which makes each detection traceable to a rule, a trace fragment, and concrete values.

**Representation of Vulnerability Knowledge by Kleene Algebra with Tests**

The knowledge base $\mathcal{R}$ represents vulnerability classes as formal rules over the state-enriched events defined above. Rules follow grammar (2). The event predicates form a finite set $\mathcal{A}$ : each $a: E \to \{0,1\}$ checks whether an event belongs to a given instruction class (for example, arithmetic events *add*, *adds*, *sub*, *subs*, *mul*; or store events *str*, *strb*, *strh*). The state predicates form a finite set $\Phi$: each $\varphi: S \to \{0,1\}$ checks a condition over the concrete state, the typical ones being $V = 1$, $addr = 0$, $addr >$ 0x1000, and $\neg\,\text{in_alloc}(addr)$.

The semantics of a rule $r$ is a language $\mathcal{L}(r) \subseteq E^*$ over finite trace fragments. For an event $e = (\iota, op, s)$,

$$\mathcal{L}(a) = \{(e) \in E^* : a(e) = 1\}, \qquad (7)$$

$$\mathcal{L}([\varphi]) = \{(e) \in E^* : \varphi(s) = 1\}, \qquad (8)$$

and, for compound expressions,

$$\mathcal{L}(r_1; r_2) = \{uv : u \in \mathcal{L}(r_1),\ v \in \mathcal{L}(r_2)\}, \qquad (9)$$

$$\mathcal{L}(r_1 + r_2) = \mathcal{L}(r_1) \cup \mathcal{L}(r_2), \qquad (10)$$

$$\mathcal{L}(r^*) = \bigcup_{n \geq 0} \mathcal{L}(r)^n. \qquad (11)$$

A rule therefore defines not only an instruction pattern but a set of trace fragments in which both the event sequence and the corresponding state predicates are satisfied.

The current rule base contains three CWE-oriented rules. The integer-overflow rule for CWE-190 is

$$r_{\text{CWE190}} = \text{any}^*; (add + adds + sub + subs + mul); [V = 1],$$

which fires on any add-like arithmetic instruction whose overflow flag is set. In the implementation the same rule is stored in JSON form in Fig. 2.

```
kat_v1_rules.json  •  CWE-190 rule
1  {
2    "id":   "CWE-190-signed-overflow",
3    "cwe":  "CWE-190",
4    "term": [
5      "any*",
6      {
7        "add|adds|sub|subs|mul": ["dst", "src1", "src2"],
8        "where": "%V == 1"
9      }
10   ]
11 }
```

Fig. 2. JSON Kleene Rule

The null-pointer-dereference rule for CWE-476 is

$$r_{\text{CWE476}} = \text{any}^*; (ldr + \cdots + str + \cdots); [addr = 0],$$

which fires when a load or store uses an effective address equal to zero. The heap-buffer-overflow rule for CWE-122 is

$$r_{\text{CWE122}} = \text{any}^*; (str + strb + strh); [addr > 0\text{x}1000 \wedge \neg\, \text{in_alloc}(addr)],$$

which fires when a store writes to an address above 0x1000 that is not contained in any live allocated block. These three rules constitute the knowledge base

$$\mathcal{R} = \{r_{\text{CWE190}}, r_{\text{CWE476}}, r_{\text{CWE122}}\}.$$

The same representation can describe multi-step mechanisms (Figure 3). A heap-related vulnerability, for instance, links an arithmetic result, a later allocation size, and a subsequent store address: the event predicates select the relevant instruction classes, while the tests preserve equality, range, and membership relations between values stored in machine states. This is the main advantage of Kleene algebra with tests over single-instruction checking: a rule may encode a vulnerability as a trace-level mechanism rather than an isolated event.

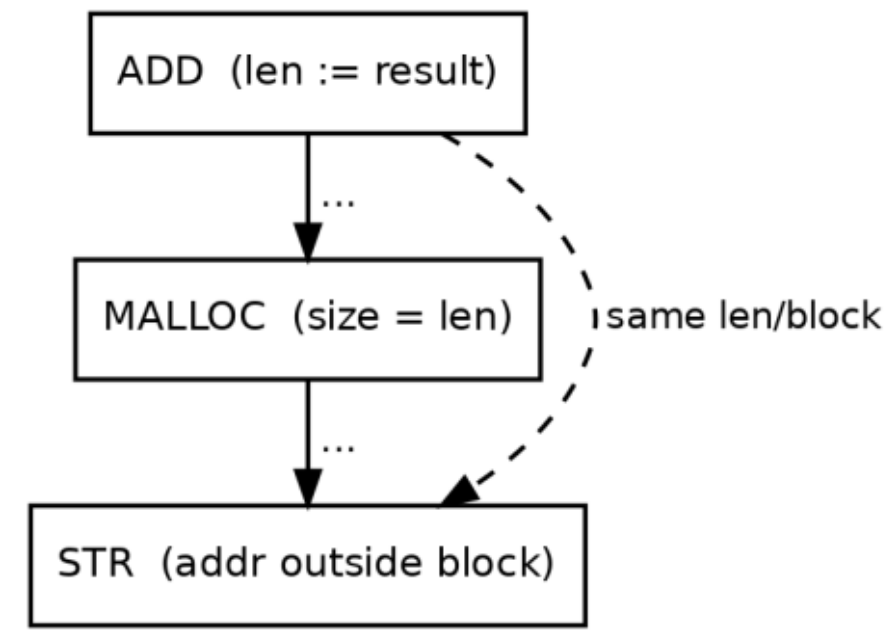

Fig. 3. A multi-step rule links one value across three events

**Lemma 1.** *Every rule $r$ generated by grammar (2) defines a regular language over the extended alphabet of events equipped with state predicates.*

**Proof.** Each event predicate $a$ denotes a set of one-event words and is regular. Each test $[\varphi]$ denotes a set of one-event words selected by a predicate over the state component, hence is regular over the extended alphabet. Regular languages are closed under concatenation, union, and Kleene iteration; since the grammar builds rules only by these operations on elementary event and test languages, $\mathcal{L}(r)$ is regular for every $r$. Therefore each rule can be compiled into a finite automaton.

Lemma 1 provides the basis for automaton-based detection: since every rule defines a regular trace language, matching reduces to finite-state recognition, and the tests require no SMT solving because they are evaluated on concrete states produced by the twin.

**Automaton-Based Inference Mechanism**

For each rule $r \in \mathcal{R}$ the inference engine uses the automaton $A_r$ of (5) as the symbolic inference mechanism that applies the rule to the trace. A transition is enabled by two conditions: the current event must satisfy the required event predicate, and the current state must satisfy the corresponding test. For an event $e_t = (\iota_t, op_t, s_t)$ a transition has the form $q \xrightarrow{a,\varphi} q'$, taken at step $t$ if $a(e_t) = 1$ and $\varphi(s_t) = 1$ ; a transition without a state condition uses the constant predicate $\varphi(s) = 1$.

The trace $\tau = (e_0, \dots, e_T)$ is processed in one pass. At each step the engine updates the active automaton states of all rules, without enumerating alternative paths and without invoking an SMT solver. For example, the CWE-190 automaton (Figure 4) stays in a scanning state while arbitrary events are read; when an arithmetic event from the specified class is observed, it checks the overflow flag in the post-instruction state, and if $V = 1$ the accepting state is reached and a CWE-190 detection is generated.

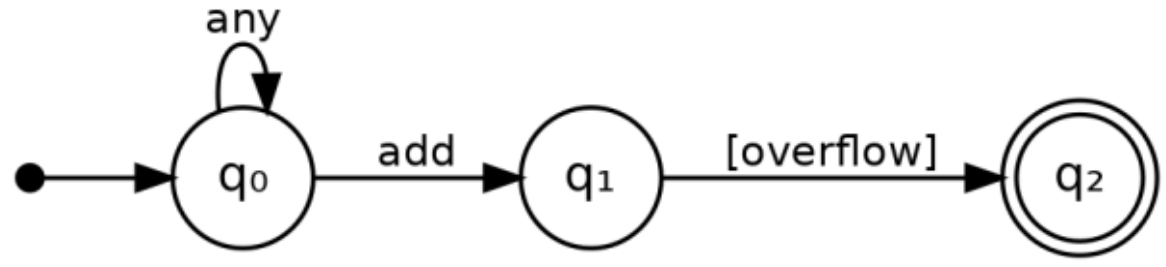

Fig. 4. Automaton for the integer-overflow rule (CWE-190)

A match for $r$ is fixed when the automaton reaches an accepting state, $q_t \in F_r$. The engine then stores the matched positions $I = (i_0, \dots, i_k)$ and the concrete values that made the transition sequence valid, and emits a detection report of the form (6). For CWE-190, $V_d$ may contain the destination value, the source operands, and $V = 1$; for CWE-476, $addr = 0$; for CWE-122, the store address and the fact that it lies outside every live block. The procedure is summarized in Algorithm 1.

**Algorithm 1. Automaton-based symbolic inference**

Input: trace $\tau = (e_0, \dots, e_T)$, rule base $\mathcal{R}$.
Output: set of detection reports $\mathcal{D}$.

1. For each rule $r \in \mathcal{R}$, construct the automaton $A_r$.
2. Initialize the active state of each $A_r$ to its initial state $q_r^0$.
3. For each event $e_t = (\iota_t, op_t, s_t)$ : update every active automaton state using the event predicate $a(e_t)$ and the state predicate $\varphi(s_t)$.
4. If an accepting state of $A_r$ is reached, create a report $d = (r, \text{CWE}, I, V_d)$.
5. Return all reports in $\mathcal{D}$.

For a fixed rule base, the cost is proportional to the trace length and the total number of automaton transitions updated. When the automata are small and the number of active states is bounded, the procedure is a linear streaming pass over the trace, which is sufficient for the compact CWE-190, CWE-476, and CWE-122 rules evaluated here.

The detector is a symbolic, knowledge-based component rather than a learned one. Vulnerability classes are stored as explicit, inspectable rules, and a positive detection is a reproducible derivation: the automaton path identifies the matched event sequence and the tests identify the concrete state facts that triggered the rule. The system does not train a model or infer rules from data. A future neuro-symbolic extension could use machine learning to rank suspicious trace fragments or to propose candidate rules, while keeping the final, explainable verification rule-based.

**Experimental Setup**

The evaluation used the prototype DIGITAL_TWIN_ARM64_v2, a concrete-state digital twin of AArch64 execution with 180 instruction meanings loaded from arm64_semantics.json. The vulnerability rules are loaded as Kleene-algebra-with-tests constraints from kat_v1_rules.json. The experiment therefore exercises the full pipeline: instruction execution, state reconstruction, event-trace generation, KAT-based matching, and report generation. Each trace was processed by the command

```
./bin/digital_twin_arm64 \
  -s semantics/arm64_semantics.json \
  --kat-constraints \
    constraints/kat_v1_rules.json \
  <trace>.json
```

The rule base $\mathcal{R}$ contained the three rules defined above. Four execution traces were used: three contain one known vulnerability each, and one is safe. The safe trace verifies that the rule base does not fire when the state predicates are false. Trace 1 (CWE-190) adds 0x7FFFFFFF and 0x1, producing 0x80000000 with $V = 1$. Trace 2 (CWE-476) loads from address 0 . Trace 3 (CWE-122) stores to 0x200000 , outside the live block [0x100000,0x100080) . Trace 4 (safe) performs a small addition 0x10 + 0x20 with no overflow and accesses only the live block [0x800,0x880).

**Results**

The prototype detected all three vulnerable traces and produced no detection on the safe trace; the results are summarized in Table 1. For each positive detection the prototype produced a report containing the CWE identifier, the triggering trace position, and the concrete explanatory values.

Table 1. Detection results on the four traces

| Trace | Expected | Detected | Triggering condition (key values) |
|---|---|---|---|
| 1 | CWE-190 | yes | V = 1, dst = 0x80000000 |
| 2 | CWE-476 | yes | addr = 0x0 |
| 3 | CWE-122 | yes | addr = 0x200000, outside [0x100000, 0x100080) |
| 4 | safe | no | V = 0, access within [0x800, 0x880) |

In Trace 1, rule $r_{\mathrm{CWE190}}$ fired after the 32-bit addition of 0x7FFFFFFF and 0x1: the result was 0x80000000 and the overflow flag $V$ was set, so the detection was justified by the arithmetic instruction class together with the post-instruction state satisfying $V = 1$. In Trace 2, rule $r_{\mathrm{CWE476}}$ fired when the program loaded from the address held in $x0$; since $x0 = 0$, the effective address was $addr = \mathrm{0x0}$. In Trace 3, rule $r_{\mathrm{CWE122}}$ fired on a store to $addr = \mathrm{0x200000}$, which lies outside the live block [0x100000,0x100080), so the detection was explained by the store event and the predicate $\neg\,\mathrm{in_alloc}(addr)$. Trace 4 served as a negative control: the addition $\mathrm{0x10} + \mathrm{0x20}$ did not set the overflow flag, and the memory accesses stayed within the live block [0x800,0x880), so none of the predicates $V = 1$, $addr = 0$, or $\neg\,\mathrm{in_alloc}(addr)$ held and no rule fired. In every case the detector linked the symbolic rule, the concrete machine state, and the generated explanation.

## Discussion

The evaluation indicates that compact Kleene-algebra-with-tests rules can capture multi-step weaknesses in AArch64 machine code while operating directly on concrete execution traces. Three properties follow from the design. Detection is solver-free: each test is evaluated on a concrete state produced by the twin, which keeps matching to a single linear pass over the trace. Because a rule combines an event pattern with state predicates and may reuse bound values across steps, it expresses vulnerability mechanisms that single-instruction checks cannot represent. And each detection is accompanied by the concrete values that satisfied the rule, which yields a reproducible explanation rather than an opaque score.

The approach also has clear limitations. Detection is confined to the supplied trace, so, unlike symbolic execution [6], [7], it does not explore alternative inputs. Its precision depends on the fidelity of the instruction semantics and on the quality of the traces. The present rule base covers three CWE classes; the safe trace produced no false positive on this controlled set, but a larger and more diverse benchmark is required to estimate precision and recall. Finally, the allocation model is a flat table of live blocks, which is adequate for the reported experiments but weaker than a full symbolic-heap model.

## Conclusions

This paper proposed an explainable Symbolic AI digital twin for vulnerability detection in AArch64 machine-code traces. The approach combines a concrete-state execution twin, a symbolic knowledge base of CWE-oriented rules in Kleene algebra with tests, and an automaton-based inference mechanism. Its methodological core is the reduction of binary-level vulnerability detection to trace pattern recognition under state-dependent predicates: unlike SMT-based symbolic execution, the mechanism does not solve path constraints during detection but evaluates explicit predicates over concrete states, which makes each detection directly described through a rule, a trace position, and concrete evidence such as $V = 1$, $addr = 0$, or an address outside a live block.

The experimental validation confirmed the expected behaviour on four controlled traces: the system detected integer overflow (CWE-190), null pointer dereference (CWE-476), and heap buffer overflow (CWE-122), and produced no detection on the safe trace. The current implementation is limited to concrete traces, 180 AArch64 instruction meanings, and a small rule base. Future work will extend the rule set to further CWE classes, including use-after-free (CWE-416) and time-of-check to time-of-use (CWE-367), improve the memory model, support larger traces, and evaluate the approach on real firmware and binary datasets.